# DNA Nanotechnology for Superradiance

Jaewon Lee[1,+], Sung Hun Park[1,+], Jangwon Kim[1,+], Kyung Hun Rho[1], Hoyoung Lee[1], Soyeon Kim[1], Seungwoo Lee[1,2*]

[1] KU-KIST Graduate School of Converging Science and Technology, Korea University, Seoul 02841, Republic of Korea

[2] Department of Integrated Energy Engineering (College of Engineering) and Department of Biomicrosystem Technology, Korea University, Seoul 02841, Republic of Korea

* seungwoo@korea.ac.kr

+ equally contributed

Superradiance, first proposed by Dicke in 1954, is a highly efficient quantum light source that differs from conventional spontaneous emission. Unlike typical spontaneous emission, where intensity scales linearly with the number of electric dipoles, superradiance exhibits an intensity that scales quadratically with the number of electric dipoles. Similarly, the decay rate also increases proportionally to the dipole numbers. This collective emission is especially powerful when it manifests as superfluorescence, a physical regime of superradiance where spontaneously emerging coherence is achieved by arranging excited dipoles in the same orientation within a volume much smaller than their emission wavelength. Numerous experimental strategies have been employed to generate superradiance, with one common approach being the use of stochastically formed aggregates of quantum dots and organic dyes. However, the inherent randomness in such systems prevents precise control over the number, spatial distribution, and relative orientation of the emitters. This often leads to non-uniform coupling strengths and parasitic dephasing effects, which make it difficult to predict the resulting quantum emission and limit its use in engineered devices. A deterministic platform that provides precise control over these parameters is therefore essential for realizing the full potential of superradiant systems. Here, we (i) specifically outline the advantages of DNA nanotechnology in tackling this challenge, (ii) discuss the reasons why superradiance has not yet been realized even with the state-of-the art DNA nanotechnology, and (iii) propose potential solutions for overcoming the current limitations.

## 1.1 Superradiance and molecular engineering

In 1954, R. H. Dicke first introduced the concept of superradiance (Dicke 1954). When the number of atoms or molecules ($N$), excited by the illumination of a general light field, form dipoles and are arranged at intervals much smaller than the wavelength of this light (Fig. 1-1), they can interact in a collective and coherent manner. As a result, they can generate superradiance, a phenomenon similar to lasing. In this seminal paper, Dicke describes how optical dipoles within a closely packed cluster interact with one another through their shared electromagnetic field and radiate coherently (Fig. 1-1). In the ideal case, this phenomenon causes a cluster of $N$ dipoles, excited coherently, to emit a pulse whose intensity and decay rate are enhanced by factors of $N^2$ and $N$, respectively, compared to an isolated dipole (Cong 2016, Gross 1982).

The key to achieving a deterministic implementation of superradiance lies in the ability to precisely assemble a desired $N$ of atoms or molecules—capable of forming electric dipoles upon excitation by light—in specific orientations and at exact intervals. Micro/nanofabrication techniques, including semiconductor processing, have inherent limitations, particularly in simplifying complex structures at the atomic or molecular scale and achieving seamless 3D integration. Even state-of-the-art methods such as extreme ultraviolet (EUV) lithography face these constraints. These challenges highlight the potential of self-assembly as a complementary approach.

Indeed, experimental observations of superradiance have been reported using random aggregates of quantum dots (Fig. 1-2) and organic dyes, such as J- or H-aggregates (Fig. 1-2) (Scheibner 2007, Bricks 2018, De Boer 1990, Fidder 1990, Spano 2000, Meinardi 2003). More recently, a nanohole-array aperture has been utilized to induce phase-controlled atom-cavity interactions, leading to the observation of superradiance (Fig. 1-2) (Kim 2018). However, no fabrication method has been reported that fully satisfies the aforementioned requirements for achieving superradiance.

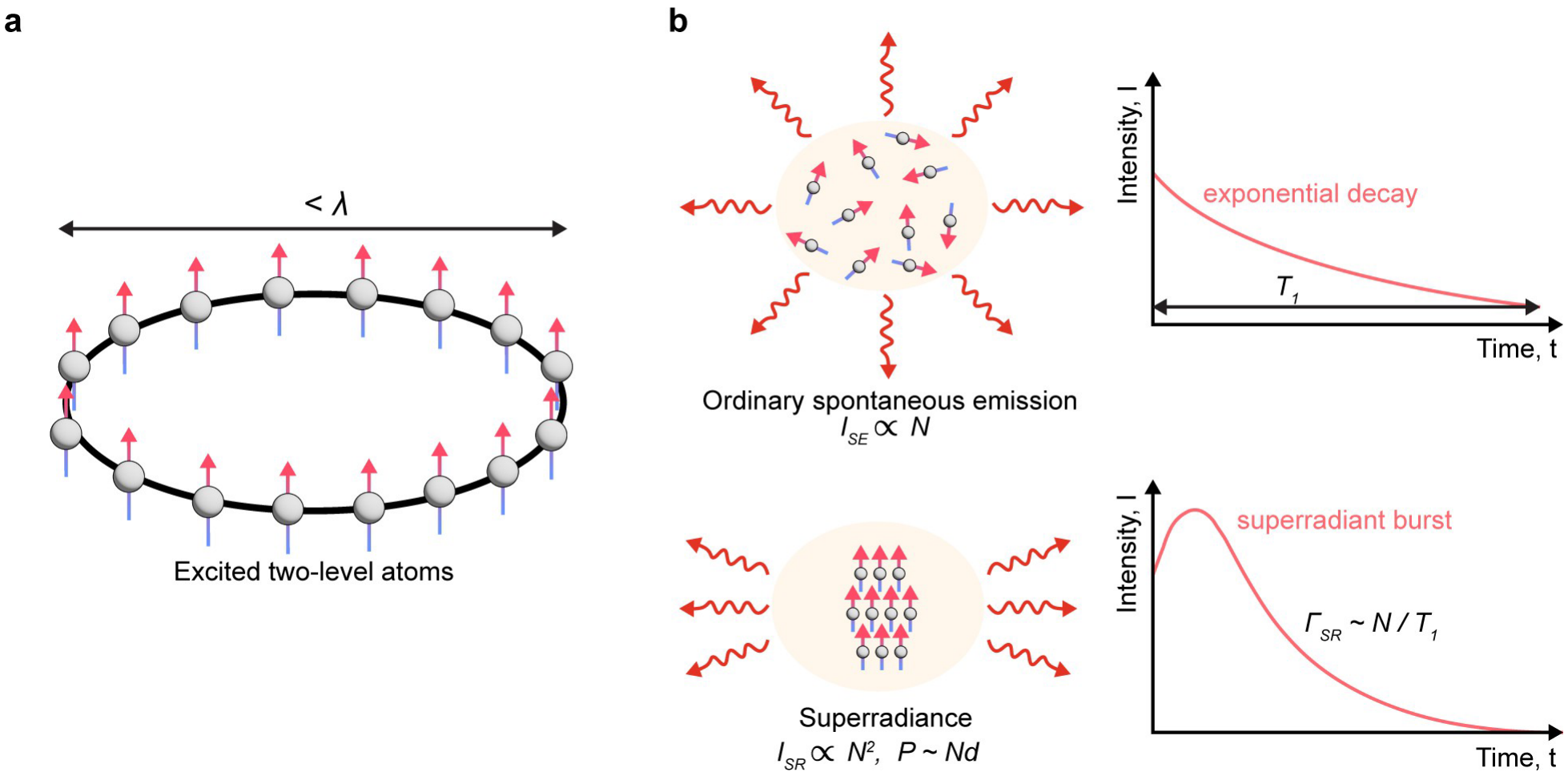

Figure 1-1: Superradiance in a collective system of atomic or molecular dipoles. (a) Schematic illustration of excited two-level atoms arranged at subwavelength intervals (<$\lambda$) under illumination. The dipoles, represented by arrows, exhibit phase locking through correlated dissipation, leading to superradiance. (b) Comparison of emission characteristics: (top) ordinary spontaneous emission from an ensemble of $N$ excited two-level atoms, and (bottom) superradiance, where coherence enhances both intensity and decay rate. $I_{SE}$, $I_{SR}$, $T_1$, $\Gamma_{SR}$, $P$, and $d$ correspond to spontaneous emission intensity, superradiant emission intensity, spontaneous radiative decay time, decay rate of superradiant emission, superradiant giant dipole moment, and individual atomic dipole moment, respectively.

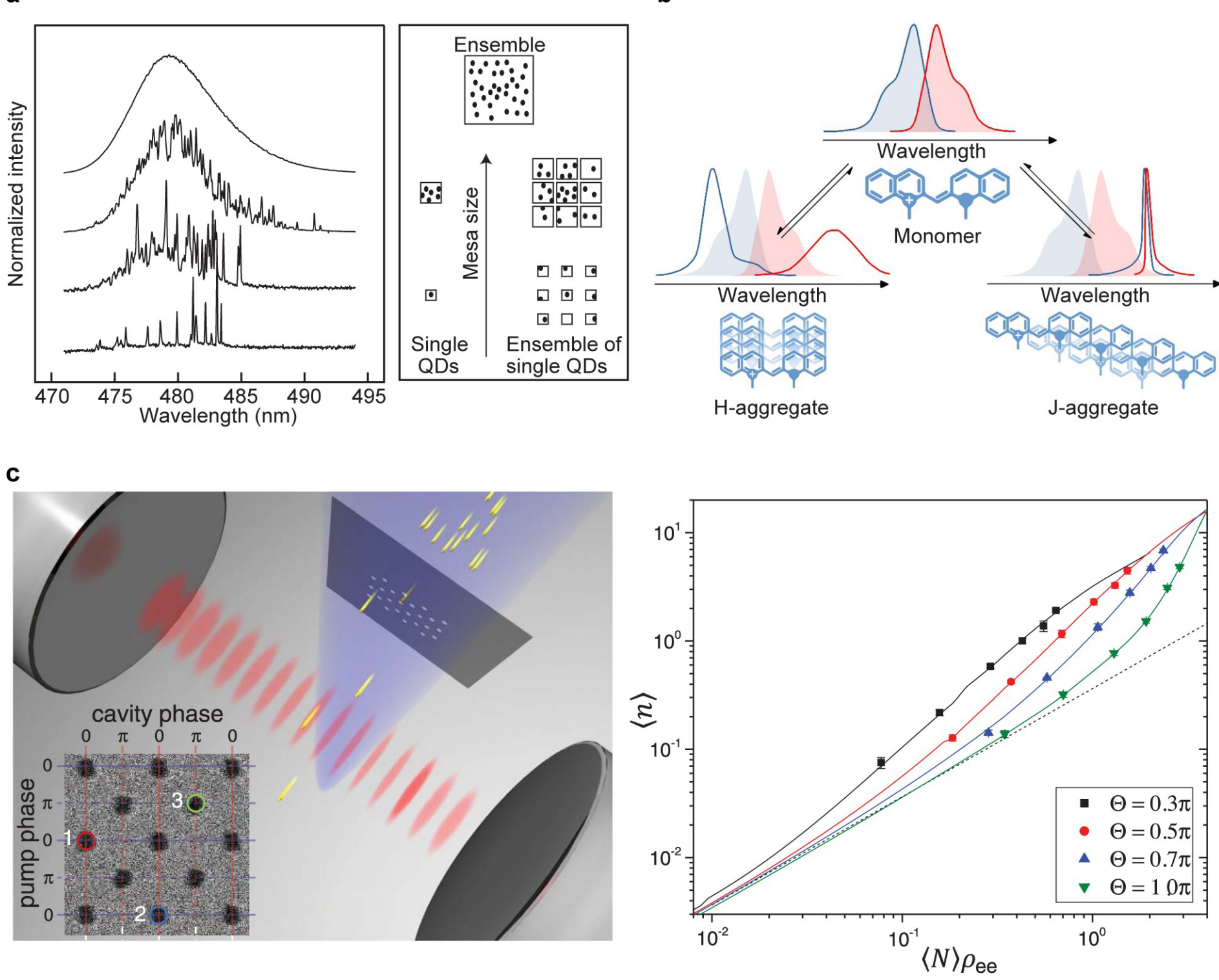

Figure 1-2: Experimental demonstrations of superradiance. (a) Photoluminescence spectra of quantum dots (QDs) placed on mesas. Arrays of the mesas facilitate the fabrication of spatially well-separated QDs exhibiting superradiant behavior. Adapted from (Scheibner 2007). Copyright 2007

Springer Nature. (b) Schematic representation of organic dye molecules self-assembled into J- or H- aggregates, which exhibit collective optical properties associated with superradiance. Adapted from (Bricks 2018). Copyright IOP Publishing. (c) Cavity-mediated coherent superradiance from a single atom. (left) Schematic of atom-cavity interactions facilitated by a nanohole-array aperture. (right) Plot of intracavity mean photon number $\langle n \rangle$ as a function of the excited-state mean atom number $\langle N \rangle \rho_{ee}$, demonstrating superradiant emission. Adapted from (Kim 2018). Copyright 2018 American Association for the Advancement of Science.

## 1.2 Why DNA?

Structural DNA nanotechnology offers a promising pathway to atomic- and molecular- scale precision, providing an exciting opportunity to bridge molecular-scale features with micron- and nanoscale structures. In particular, the development of DNA origami has accelerated these technological advancements as follows.

B-form DNA double helices have a width of approximately 2.0–2.5 nm, with a base pair rotation angle of 33–34 degrees and an inter-base pair distance of 0.33–0.34 nm (Fig. 1-3). DNA origami can be viewed as a molecular ”raft,” where such standardized DNA double helices are joined together in a controlled manner, much like wooden logs tied together to form a floating ship. This assembly is made possible by the fundamental property of DNA—Watson and Crick’s complementary base pairing (Fig. 1-3). By designing specific sequences, the length and position of the DNA double helices to be jointly connected together can be precisely controlled at the molecular level, enabling the fabrication of nearly any shaped 2D or 3D DNA nanostructures (Fig. 1-3) (Rothemund 2006, Douglas 2009).

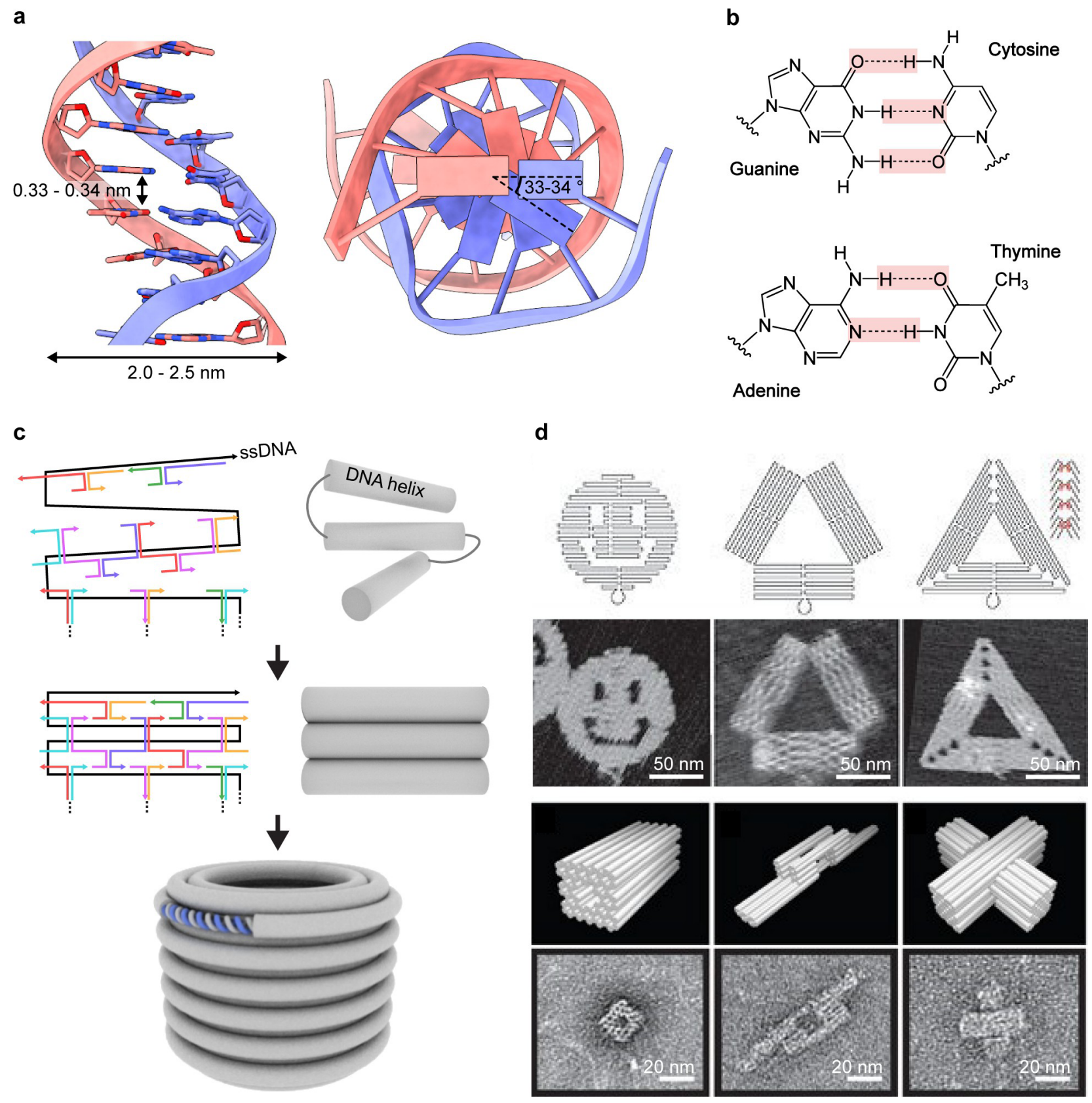

Figure 1-3: Fundamental principles of DNA origami assembly. (a) Schematic representation of the B-form DNA double helix, shown in front (left) and top (right) views. (b) The four nucleotide bases and their complementary bindings. (c) Schematic illustration of DNA nanostructure assembly through sequence-programmed hybridization. (d) Examples of 2D and 3D DNA origami structures. Adapted from (Rothemund 2006, Douglas 2009). Copyright 2006 Springer Nature. Copyright 2009 Springer Nature.

More importantly, DNA origami can be used as a molecular pegboard, to arrange quantum dots or organic dyes (Cho 2022). The most widely used approach involves decorating the DNA origami surface with single-stranded DNA (ssDNA) handles, which serve as anchor points. These handles can hybridize with complementary ssDNA strands, known as anti-handles, that are attached onto quantum dots or organic dyes, thereby enabling programmable assembly of quantum dots and organic dyes on the DNA origami surface (Fig. 1-4). The number of handle ssDNA molecules can be digitally controlled and their spatial positioning can be achieved at molecular precision (generally, known spatial resolution is about few nanometers). These arrangements of handle ssDNA on DNA origami can be quantitatively verified using DNA-PAINT, a super-resolution optical microscopy technique (Schnitzbauer 2017).

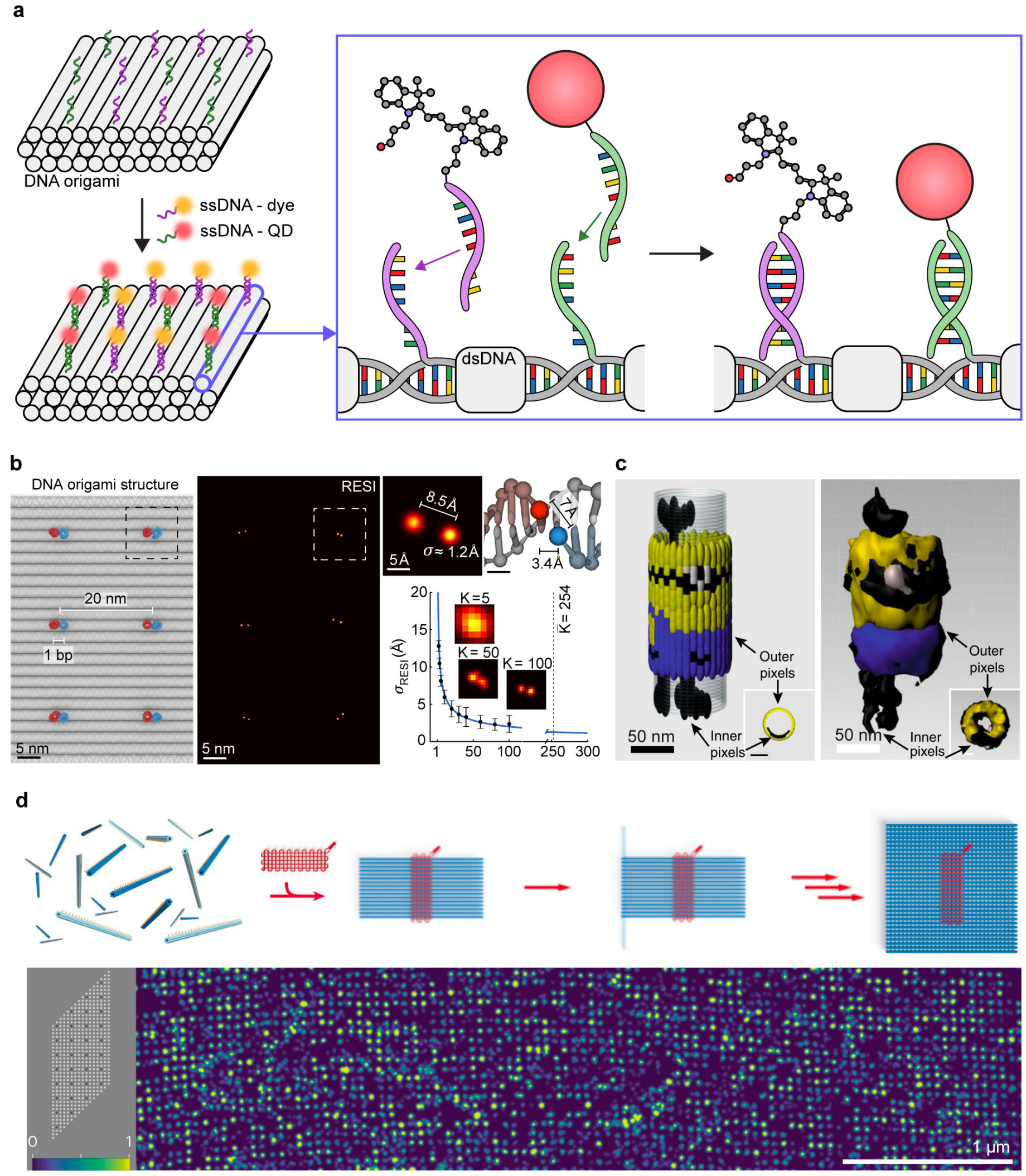

Figure 1-4: DNA origami as a molecular pegboard. (a) Schematic representation of DNA origami with single-stranded DNA (ssDNA) handles for site-specific attachment of target molecules. (b) Super-resolution DNA-PAINT imaging resolving angstrom-level spacing. Adapted from (Reinhardt 2023). Copyright 2023 Springer Nature. (c) Model of a co-axially stacked DNA origami barrel (left) and corresponding DNA-PAINT image (right). Adapted from (Wickham 2020). Copyright 2020 Springer Nature. (d) Crisscross polymerized DNA origami megastructure as a multi-micron scale molecular pegboard. (left) Schematic illustration of the crisscross assembly of DNA origami. (right) DNA-PAINT image of the periodic crisscross megastructure. Adapted from (Wintersinger 2023). Copy- right 2023 Springer Nature.

Recent advances in DNA-PAINT have demonstrated the ability to resolve organic dye spacings even at the angstrom level (Fig. 1-4) (Reinhardt 2023). While DNA-PAINT quantifications of the spatial arrangement and spacing resolution of the clustered dyes have been traditionally performed on 2D DNA origami pegboards, recent studies have reported the feasibility of 3D integration of dyes using co-axially stacked DNA origami barrels (similiar to Dicke's early dream) (Fig. 1-4) (Wickham 2020). In addition, the lateral dimension of individual DNA origami, which was previously limited to around 100 nm–or a few hundred nanometers when multiple DNA origami units were linked together–has now been significantly expanded to ~2 μm using a novel hierarchical joining strategy known as crisscross polymerization, which can effectively suppress spurious nucleation of DNA origami and promote the all-or-nothing crystallization of DNA origami unit (slats) until the desired micron-scale is achieved (Wintersinger 2023). This implies the possibility of scalable integration of superradiant sources by extending arrays of organic dyes (or quantum dots) to the micrometer scale (Fig. 1-4).

Furthermore, DNA origami can be positioned with extremely high precision on integrated photonic chips, a technique known as DNA origami placement (DOP) (Fig. 1-5) (Gopinath 2016, Martynenko 2023, Gopinath 2021). The DNA origami surface is negatively charged and surrounded by positive ions such as $Mg^{2+}$. Consequently, DNA origami can adsorb onto a negatively charged chemical patch where these $Mg^{2+}$ ions can selectively bind ($Mg^{2+}$ bridges the DNA origami and chemical patch). When this chemical patch retains shape and size complementarity with the DNA origami, it not only controls the spatial position of the DNA origami but also guides its orientation, enabling directed organization of DNA origami onto the chemical patch-arrayed chip (DOP). Similar to general equilibrium self-assembly, when the interaction between the chemical patch and DNA origami is neither too strong nor too weak, the DNA origami undergoes repeated binding and unbinding processes with chemical patches until reaching self-alignment (minimizing enthalpy).

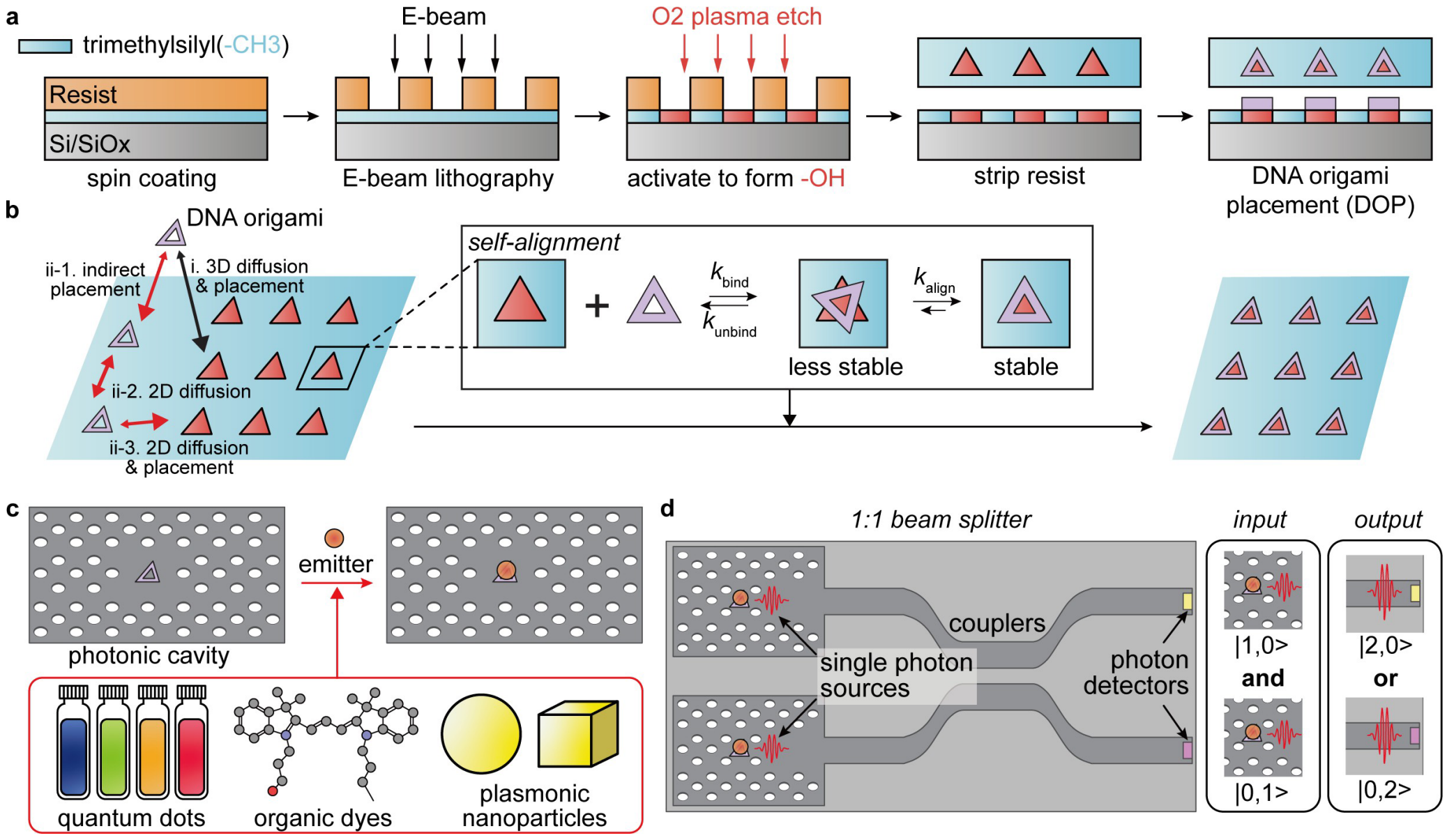

Figure 1-5: Molecular integration on photonic chips through DNA origami placement (DOP) and its potential application in photonic and quantum systems. (a) Schematic representation of the DOP process. Arrays of binding sites are fabricated by electron-beam lithography. An $O_2$ plasma can etch the residual resist and activates a trimethylsilyl layer, generating surface silanol. After lift-off, DNA origami is placed onto the chemical patches. (b) Mechanism of DOP. DNA origami attaches to the chemical patches via two routes: (i) direct binding through 3D diffusion (black arrow) and (ii) indirect binding through background binding and 2D diffusion (red arrows). Then, the DNA origami on the patch undergoes a self-alignment process, where transient binding and unbinding minimize surface energy. When the overlap region reaches approximately 100%, the DNA origami is stably fixed in a specific position and orientation. (c) Molecular integration on photonic cavity via DOP. The placed DNA origami on photonic cavity can host emitters such as quantum dots, organic dyes, and plasmonic nanoparticles. (d) Schematic representation of Hong-Ou-Mandel (HOM) model with independently tuned single photon sources that interfere at a 1:1 beam splitter followed by integrated photon detectors.

DOP could be particularly pivotal for the following reason. Generally, the development of integrated photonic platforms is essential for the immediately practical usage of quantum information processing and sensing. While the specific requirements may vary depending on the intended application, certain universal criteria remain fundamental, including (i) the deterministic generation of single or entangled photons and (ii) efficient nanophotonic elements for routing and manipulating photons. The first requirement can be addressed by integrating quantum light sources capable of superradiance onto an integrated photonic chip, while the second can be fulfilled by coupling these quantum light sources with resonant cavities. In this context, if superradiance can be realized on a DNA origami platform, DOP would enable the simultaneous satisfaction of both requirements. Notably, DNA origami serves as a versatile molecular pegboard not only for arranging quantum dots and organic dyes but also for organizing chemically synthesized, atomically smooth single-crystal plasmonic nanoparticles (Fig. 1-5). This capability further extends its potential to the integration of plasmonic cavities. Aforementioned advancements in large-scale DNA origami assembly and its deterministic placement on photonic chips elucidate its potential as a powerful tool for integrating molecular-scale, superradiant components into functional photonic and quantum systems, potentially enabling, for example, a new type of on-chip Hong-Ou-Mandel (HOM) model (Fig. 1-5) (Hong 1987).

## 1.3 Limitations of DNA

Despite the promising advantages of DNA origami-based structural DNA nanotechnology, the strong recognition of its potential for realizing superradiance, and significant research efforts dedicated to this goal, no successful demonstration of DNA nanotechnology-based superradiance has been achieved thus far. This is due to the compounded limitations of the structural DNA nanotechnology in terms of the realizing superradiance, as detailed below.

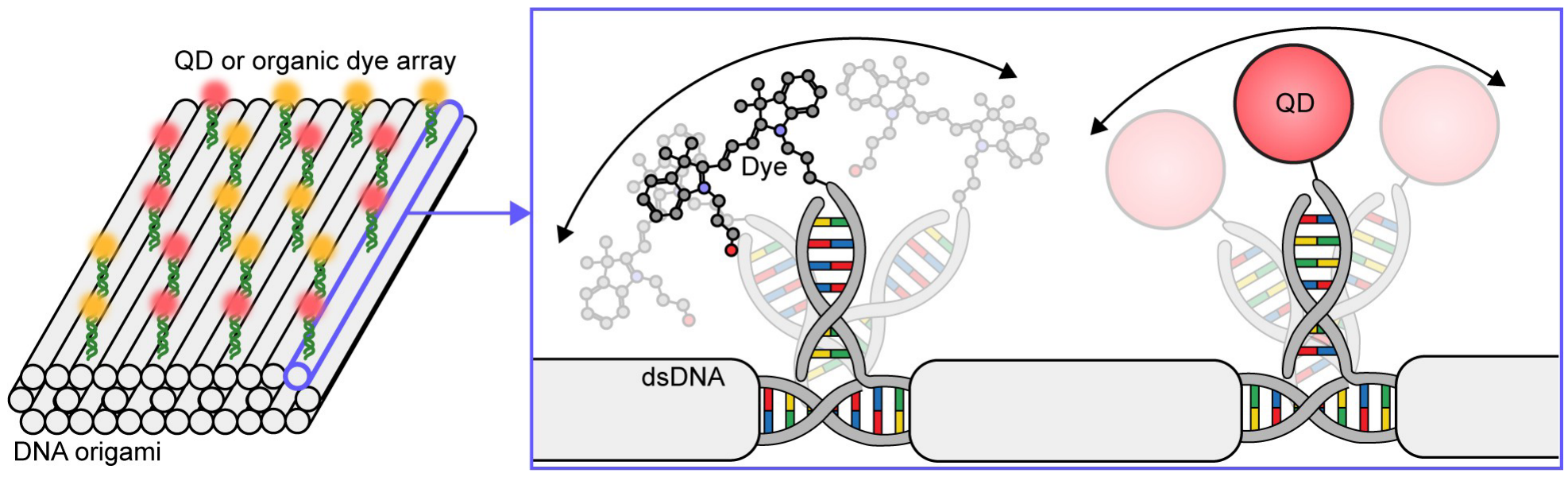

Figure 1-6: Schematic illustration of the uncontrolled pivoting motion of the handle-anti-handle DNA duplex on a DNA origami surface. Target molecules or materials tethered via this duplex experience dynamic fluctuations in solution, resulting in random orientations

The primary challenge originates from the intrinsic limitations of the widely used method of integrating quantum dots and organic dyes onto DNA origami using handle and anti-handle complementary binding. While this approach enables the precise molecular-level arrangement of a digitally controlled number of quantum dots and organic dyes, it does not allow for control over their orientations. Specifically, quantum dots and organic dyes linked to the surface of DNA origami via “DNA duplex linker” tend to ”hang out,” much like dangling bonds in surface chemistry. When self-assembled in solution, the handle and anti-handle DNA duplexes inevitably undergo pivoting motions (Fig. 1-6). In other words, since the DNA duplexes and the attached quantum dots or organic dyes are in direct contact with solvent molecules, thermal fluctuations of the solvent molecules can transfer their collision energy to these quantum dots or organic dyes, leading to unavoidable thermalization effects such as pivoting motions. As a result, quantum dots and organic dyes integrated onto DNA origami exhibit significant orientational distributions (Pan 2014). Even if these DNA origami-quantum dot/organic dye assemblies could be positioned onto “solid-state” photonic chips via DOP, the heat, that can be possibly generated during light-field exposure for superradiance characterization, could cause structural stability issues.

The latter issue could potentially be mitigated through the recent development of DNA origami silicification (Liu 2018, Nguyen 2018), which not only fixes quantum dots and organic dyes onto the DNA origami surface but also leverages the phononic vibrations of silica to facilitate heat dissipation. However, since silicification is performed after the quantum dots and organic dyes already have a broad orientational distribution, it inevitably results in the fixation of pre-existing misalignments, thereby compromising the overall spatial precision of the array.

Although alternative approaches have been proposed, such as fixing dyes at the nicks of DNA duplexes or selectively removing base pairs from one strand of a duplex to achieve finer control over their long-axis orientation (Fig. 1-7) (Cannon 2017, Adamczyk 2022), these methods present additional challenges. Not only is it difficult to precisely control the number of orientation-defined organic dyes being integrated, but more critically, they offer limited flexibility in controlling inter-dye spacing with high degrees of freedom.

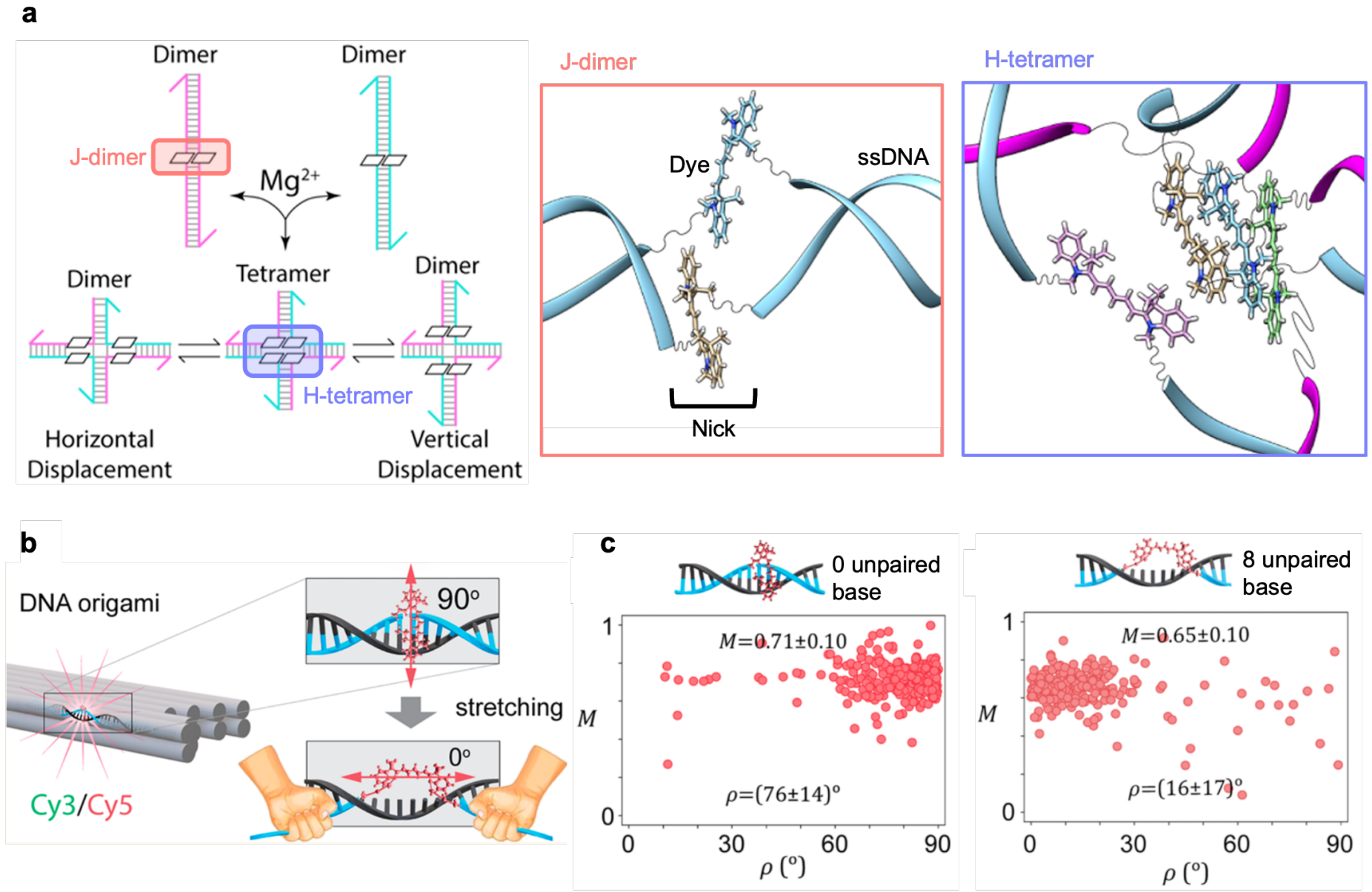

Figure 1-7: Strategies for controlling dye molecule orientation in DNA duplexes. (a) Orientation control via internal functionalization of dye molecules at DNA duplex nicks. At low salt concentrations, dyes exhibit J-aggregate behavior, whereas increased salt concentrations trigger branch migration, leading to tetramer formation with H-aggregate behavior. Adapted from (Cannon 2017). Copyright 2017 American Chemical Society. (b-c) Tuning dye molecule orientation by hybridizing fluorophore-labeled ssDNA staples onto DNA origami with a controlled number of unpaired bases. (b) Schematic representation of the approach. (c) Distribution of modulation, $M$ (wobbling of the orientation) and in plane angle, $\rho$ of dye molecules for different numbers of unpaired bases. Adapted from (Adamczyk 2022). Copyright 2022 American Chemical Society.

## 1.4 DNA-protein hybrids

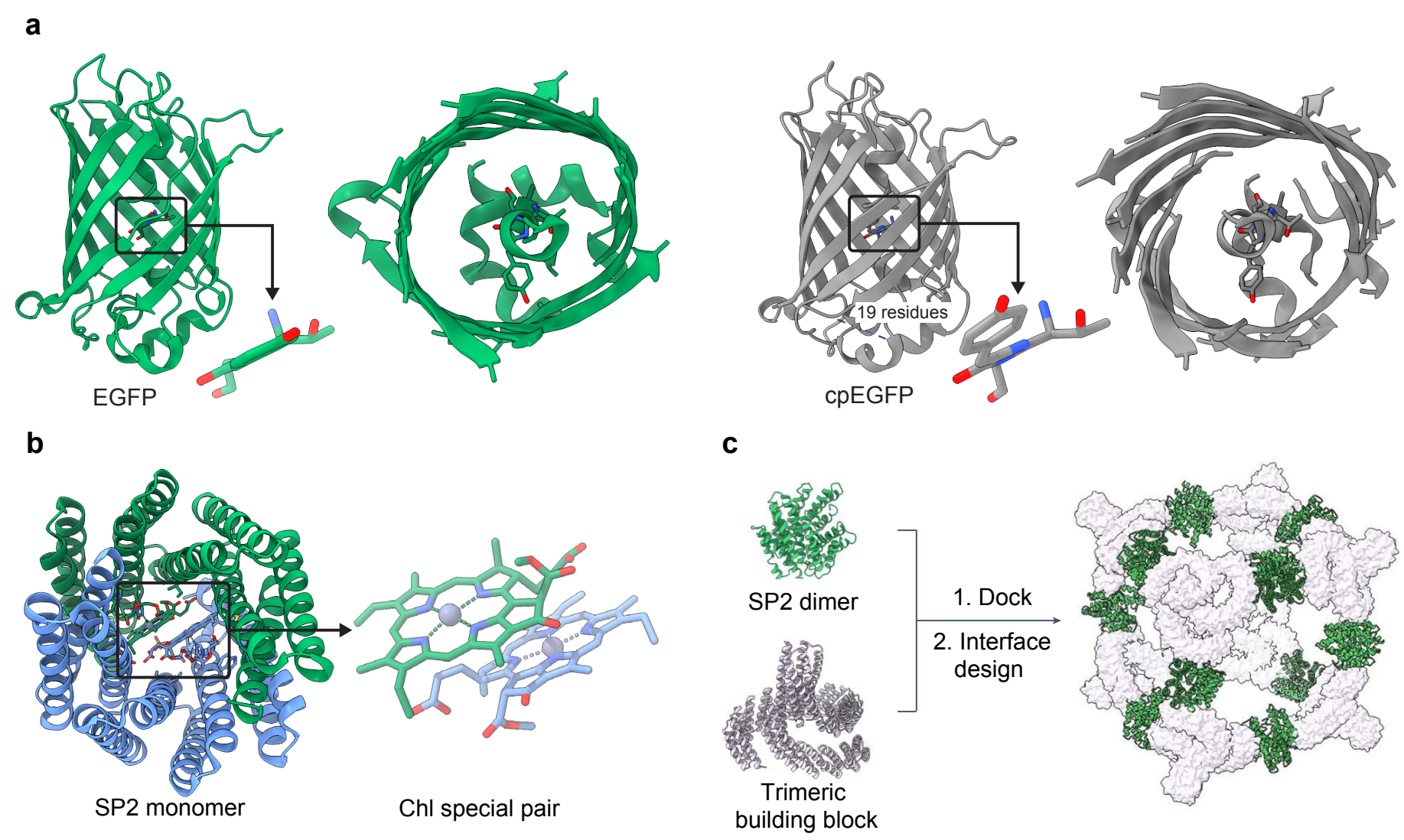

Figure 1-8: Protein scaffolds for controlling dye molecule location and orientation. (a) X-ray crystal structures of enhanced green fluorescent protein EGFP, PDB 2Y0G) (left) and circular-permutated EGFP (cpEGFP, PDB 3EVP) (right), demonstrating how chromophore positioning and orientation

can be controlled by modifying the protein scaffold. (b) X-ray crystal structure of the artificially designed special pair 2 (SP2) protein (PDB 7UNI), which precisely arranges two chlorophyll (Chl) molecules with controlled orientation and position. (c) Cryo-electron microscopy (cryo-EM) structure of SP2 protein dimers assembled into an octahedral supercomplex using trimeric cyclic oligomer scaffolds as co-building blocks. Adapted from (Ennist 2024). Copyright 2024 Springer Nature.

In addition to our arguments in the previous section, we would add one more point to that, which is it's not just the position and orientation, it's their environment as well. To realize superradiance, it is desired to get a dye that's not touching anything other than water or solvent. It's difficult to get a dye that's fixed in a vacuum or even in water because there's almost always something nearby that it's interacting with. However, this problem can be overcome by benefitting from the recent advances in protein de-novo engineering.

A particularly compelling prospect is the integration of amino acids within DNA origami frameworks. While DNA origami excels at bridging atomic-, nano- and micron-scale structures, proteins remain unparalleled in their chemical diversity and spatial precision. The ability to position molecular components with angstrom-level accuracy – especially through the use of engineered proteins – could be crucial for applications requiring the precise spatial arrangement of organic dyes (e.g., fluorophores) and other functional groups.

For instance, it has been demonstrated that a single chromophore within green fluorescent protein (GFP) can be positioned and oriented with angstrom-level precision within the protein scaffold (Fig. 1-8). Additionally, the encapsulation effect of the protein effectively shields the chromophore from interactions with external molecules, minimizing environmental perturbations.

Recent advancements in non-standard amino acids and machine learning-driven protein engineering further enhance the feasibility of such integrations, although challenges remain in fully incorporating these developments into existing computational models. For example, David Baker—recipient of the 2024 Nobel Prize in Chemistry—and his colleagues have demonstrated that de novo protein engineer- ing, guided by computational design, can achieve these structural and functional constraints artificially (Fig. 1-8) (Ennist 2024). These findings suggest that, at least in terms of digitally controlled dye orientation and positioning, as well as shielding from external molecular interference, protein engineering may offer advantages over structural DNA nanotechnology.

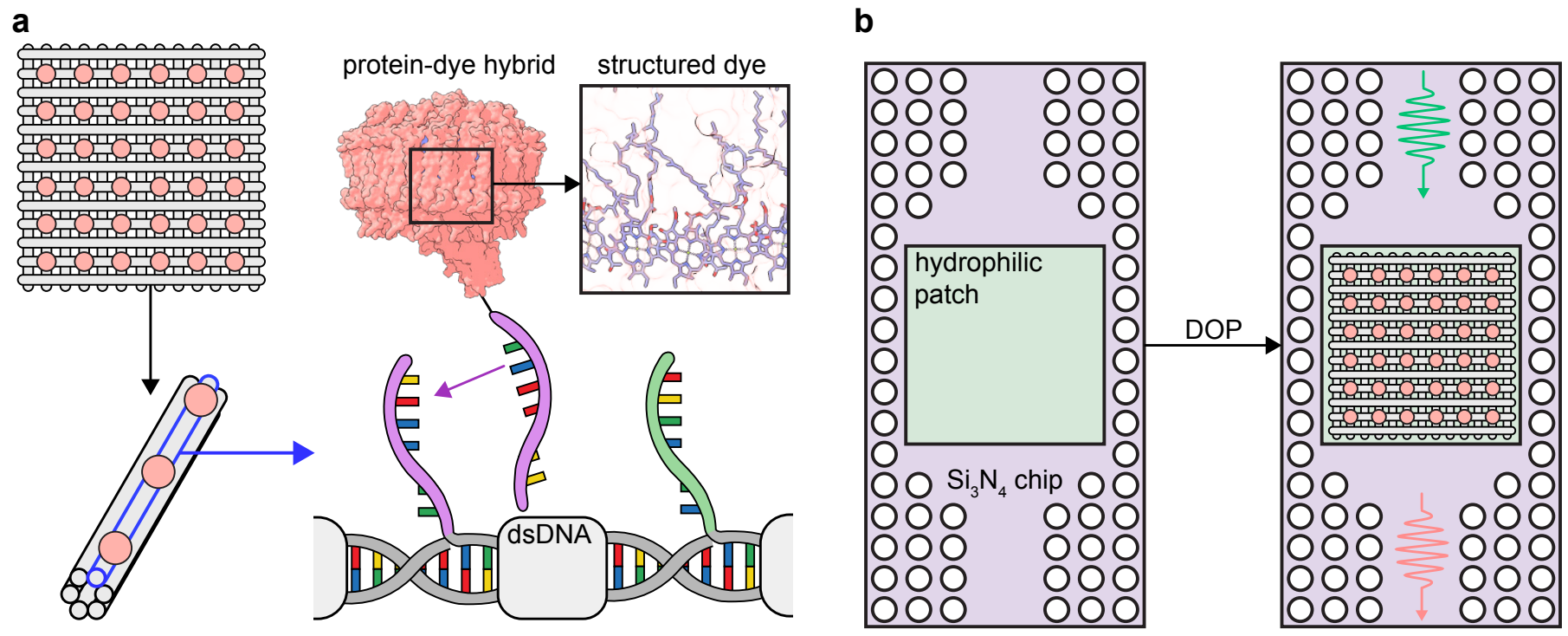

Figure 1-9: Schematic representation of DNA-protein hybrid systems for superradiance: (a) assembly of protein-dye hybrid superradiant structures on a DNA origami pegboard, and (b) integration of DNA origami-superradiant units onto photonic chips via DNA origami placement (DOP).

However, the programmable integration of de novo-engineered protein-dye hybrid structures onto photonic chips—similar to DOP—remains a challenge. This limitation suggests that for on-chip realization of superradiance and its applications in quantum processing and sensing, a synergistic interplay between protein-dye hybrids and DNA nanotechnology is necessary. One possible approach is to use de novo protein engineering to cluster a digitally controlled number of dyes within a protein scaffold, ensuring angstrom-level precision in their position, spacing, and orientation—thereby forming a single protein-dye hybrid superradiant system. To facilitate integration with DNA nanotechnology, ssDNA functioning as anti-handles can be attached to the protein exterior. These protein-dye hybrid superradiant structures can then be precisely arranged onto DNA origami surfaces via complementary binding of ssDNA strands (Fig. 1-9). Finally, these DNA origami-superradiant units can be integrated onto photonic chips using DOP, enabling the realization of programmable on-chip superradiance (Fig. 1-9).

### 1.5 Concluding Remarks

In this chapter, we discussed DNA-based molecular engineering techniques for the realization of superradiance. DNA enables the arrangement of quantum emitters, such as organic dyes, on or within the DNA double helix with nanometer-scale precision, owing to the Watson and Crick complementary base pairing. Furthermore, the orientation of these quantum emitters can be finely controlled through the selective removal or addition of base pairs. The rapid advancements in DNA nanotechnology, particularly DNA origami, have further expanded the ability to position and orient quantum emitters with high degrees of freedom in two- and three-dimensional spaces. Nevertheless, the realization of superradiance has not yet been achieved due to the intrinsic molecular vibrations. It is anticipated that this challenge could be overcome in the future through hybridization with protein engineering.

## Acknowledgement

This research was supported by a grant of the Korea-US Collaborative Research Fund(KUCRF), funded by the Ministry of Science and ICT and Ministry of Health & Welfare, Republic of Korea (grant number: RS-2024-00468463)